\documentclass[prl,twocolumn,aps,letterpaper, preprintnumbers,superscriptaddress]{revtex4}
\usepackage{hyperref}
\usepackage{verbatim}
\usepackage{amsmath}
\usepackage{latexsym}
\usepackage{revsymb}
\usepackage{yfonts}
\usepackage{ifthen}
\usepackage{tcolorbox}
\usepackage{natbib}
\usepackage{amsfonts}
\usepackage{amsmath}
\usepackage{amssymb}
\usepackage{amsthm}
\usepackage{graphicx}
\usepackage{bm}
\usepackage{bbm}
\usepackage{epsfig,color,amssymb}
\usepackage{subfigure}
\usepackage{amsfonts}
\usepackage{amscd}
\usepackage{amsmath}
\usepackage{multirow}
\usepackage{chemarrow}
\usepackage{dcolumn}
\usepackage{bm}
\usepackage{graphicx}
\usepackage{enumerate}
\usepackage{epsfig}
\usepackage{subfigure}
\usepackage{xcolor}
\usepackage{multirow}
\usepackage{ulem}
\usepackage{braket}
\usepackage{comment}
\usepackage{enumitem}
\usepackage{amsthm}

\mathchardef\myphi="0127
\mathchardef\myomega="0121
\mathchardef\myxi="0118
\renewcommand{\emph}[1]{\textit{#1}}

\begin{document}

\title{Frequency-astigmatism asymmetric nonlinear conversion of structured light lasers}

\author{Jing Pan}
\author{Hao Wang}
\author{Zijian Shi}
\affiliation{Key Laboratory of Photonic Control Technology (Tsinghua University), Ministry of Education, Beijing 100084, China}
\affiliation{State Key Laboratory of Precision Measurement Technology and Instruments, Department of Precision Instrument, Tsinghua University, Beijing 100084, China}
\author{Yijie Shen}\email{y.shen@soton.ac.uk}
\affiliation{Optoelectronics Research Centre, University of Southampton, Southampton SO17 1BJ, United Kingdom}
\author{Xing Fu}\email{fuxing@mail.tsinghua.edu.cn}
\author{Qiang Liu}\email{qiangliu@mail.tsinghua.edu.cn}
\affiliation{Key Laboratory of Photonic Control Technology (Tsinghua University), Ministry of Education, Beijing 100084, China}
\affiliation{State Key Laboratory of Precision Measurement Technology and Instruments, Department of Precision Instrument, Tsinghua University, Beijing 100084, China}


\begin{abstract}
Nonlinear optics of structured light has recently delivered intriguing fundamental physical phenomena in light-matter interactions and advanced applications from classical imaging to quantum informatics. The mutual interaction between spin, orbital angular momentum (OAM) and wavelength is extensively studied in such cases. In this work, we go beyond only considering OAM and wavelength by taking the nonlinear frequency conversion and transverse mode astigmatism conversion as two building blocks and investigating how single modes and complicated multiplexed modes evolve after them. In particular, We found a generalized law of nonlinear conversion structured light from experiments and theories, that the converted modes are highly related to the sequence of these two blocks, obeying an inherent (non)commutative rule in which. This effect not only creates extended structured laser modes but serve as new rules in nonlinear structured light manipulation.

\end{abstract}

\setlength{\parskip}{0pt}
\maketitle
When strong light interacts with a medium, most likely new frequency components are expected to be excited with a process named nonlinear optical frequency conversion that is long-investigated for years~\cite{fejer1994nonlinear,franken1961generation}. While, only recently has the spatial structure of light in nonlinear processes come to the fore~\cite{forbes2021structured,buono2022nonlinear}. Then when the fundamental wave (FW) is endowed with orbital angular momentum (OAM) carried by vortex beam~\cite{shen2019optical,naidoo2016controlled,devlin2017arbitrary}, the nonlinear process becomes implicit concerning linear and angular momenta simultaneously. Nevertheless, it would be preferable if we can extend the optical frequencies and transverse modes as we wish because of pertinent applications in quantum information processing~\cite{hu2020efficient,erhard2020advances}, superresolution microscopy~\cite{willig2006sted,vicidomini2018sted,casacio2021quantum} and holography~\cite{fang2021high}, etc. Such notion continually motivates intense interesting advances from the very beginning of vortex beam studies~\cite{PhysRevA.56.4193}, aiming to primarily understand some fundamental open issues about vortex light's nonlinear conversion~\cite{2013PhysRevA.88.063827,Zhou:14,PhysRevLett.117.163202,gauthier2017tunable,chen2020phase,zhao2020conversion,2020PhysRevA.101.043821,tang2020harmonic,gui2021second,liu2022control,zhang2022second,pires2019optical,lou2022third,wu2022conformal,de2021beyond,wu2020radial,hancock2021second}, which mainly are related with or based on FW modes with a singularity, such as the effects of phase-matching on OAM transfer from different nonlinear processes~\cite{chen2020phase}, spin-orbit interaction in second harmonic generation (SHG) process~\cite{tang2020harmonic}, and the SHG of spatiotemporal vortices carrying transverse OAM~\cite{gui2021second,hancock2021second}.

Despite these established works, the explicit nonlinear conversion rules of complex structured light with multiple controlled degrees of freedom (DoFs) beyond OAM remain elusive~\cite{he2022towards}. Furthermore, recent spatial mode shaping techniques have already tamed many typical complex structured modes, such as vortex lattices or singularity array~\cite{shen2018vortex,piccardo2022vortex}, SU(2) geometric modes~\cite{shen2020structured,shen2021rays}, optical skyrmions~\cite{shen2022generation,shen2022topological}, triggering extended applications such as high-capacity communications and encryptions~\cite{wan2022divergence,wang2022deep2,pryamikov2022rising,shen2021creation,wan2022multipartite,shen2022nonseparable,wang2022deep}, and laying the bases to explore nonlinear optics involved with general structured mode transformation. 

\begin{figure}[htbp]
\setlength{\abovecaptionskip}{0pt}
 \setlength{\belowcaptionskip}{-15pt}
 \centering
\includegraphics[width=0.8\linewidth]{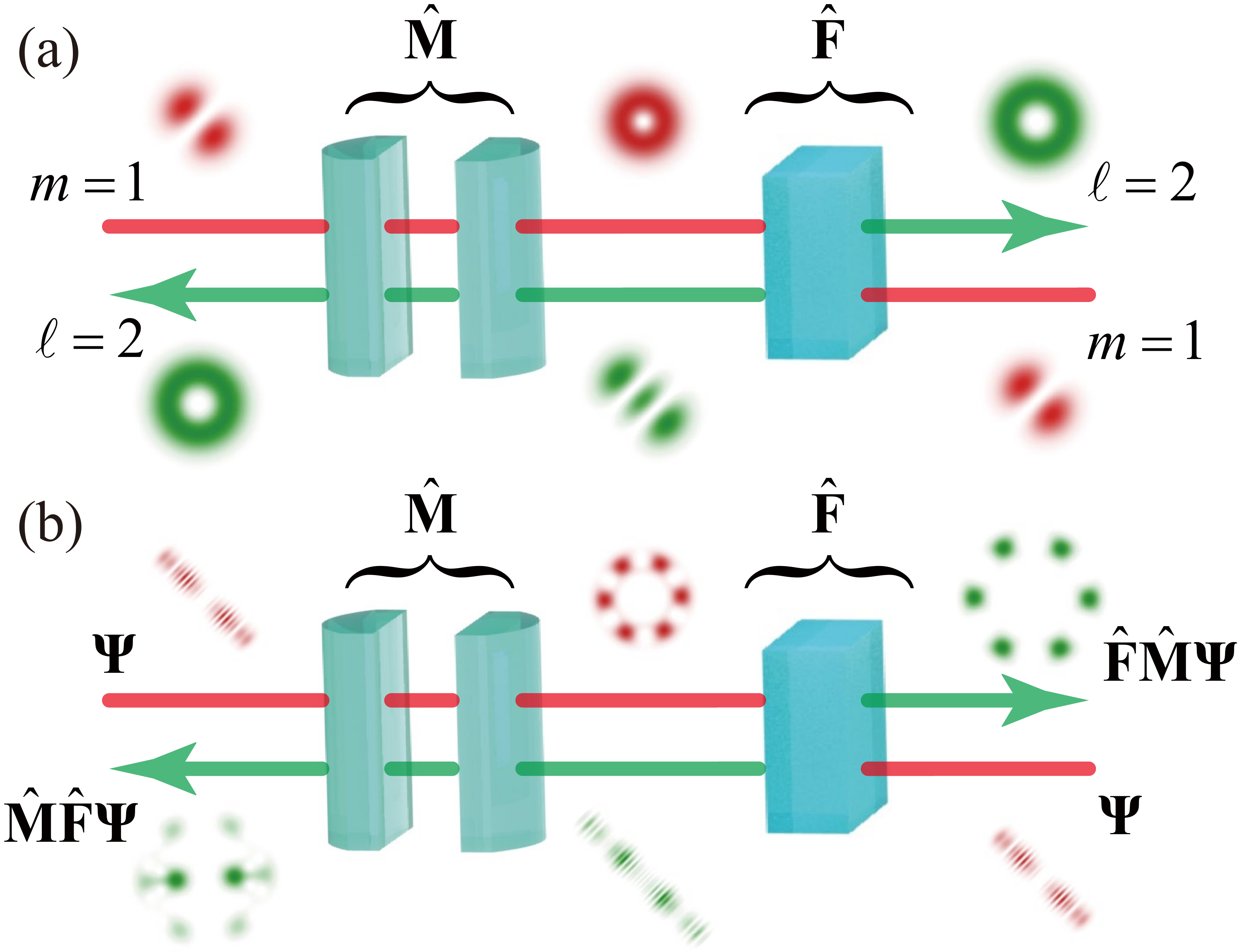}
	\caption{(\textbf{a}) Conventionally, the sequence of mode and frequency conversions ($\hat{\textbf{M}}$ and $\hat{\textbf{F}}$) does not affect the converted result. An HG mode is firstly converted into OAM mode with an azimuthal charge, then converted into SHG mode with double frequency and OAMs. The mode can also be converted into SHG mode with double frequency and topological indices, and then converted into OAM charges of two. (\textbf{b}) For the frequency-astigmatism asymmetric conversion of general structured light, the sequence of $\hat{\textbf{M}}$ and $\hat{\textbf{F}}$ is crucial, the change of the sequence will induce dramatically different results of converted topological structures and indices.} 
	\label{f1}
\end{figure}
\begin{figure*}[htbp]
 \setlength{\abovecaptionskip}{0pt}
 \setlength{\belowcaptionskip}{-15pt}
\centering\includegraphics[width=0.82\linewidth]{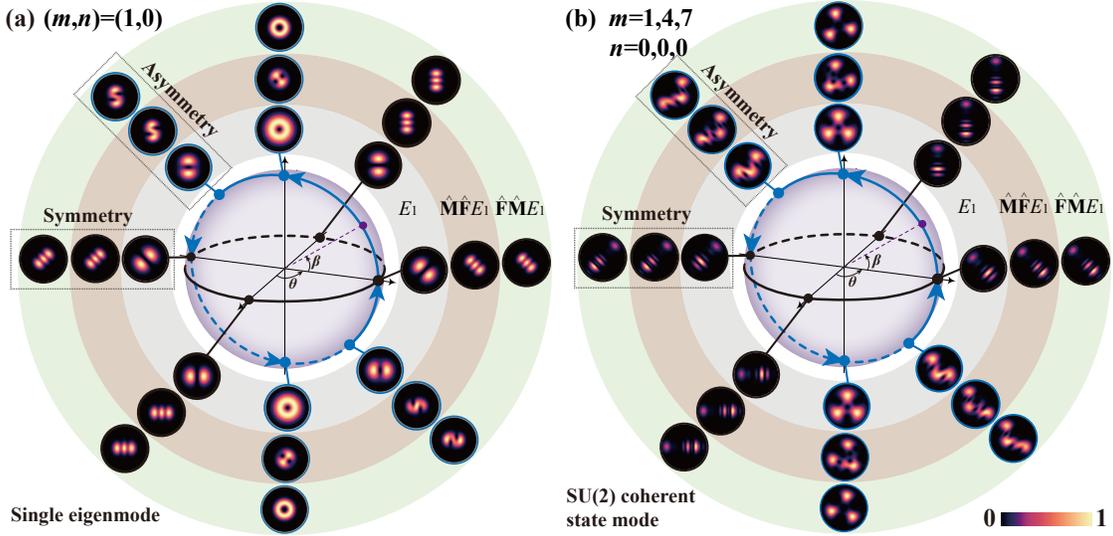}
\caption{Theoretical results of SHG combined Poincar\'e sphere of FW modes ($E_{1}$, inner gray circle), and modes after $\hat{\textbf{F}}$-$\hat{\textbf{M}}$ conversion (middle orange circles) and $\hat{\textbf{M}}$-$\hat{\textbf{F}}$ conversion (outer green circles) on (a) a fundamental HG$_{m,n}$ mode with $(m,n)=(1,0)$ and (b) an SU(2) geometric mode with 3-fold symmetry (inner gray circle), respectively. Only those points on two black equator lines represent the modes having symmetric effect of $\hat{\textbf{F}}$ and $\hat{\textbf{M}}$ conversions while other points on the spheres denote the asymmetric results.}
\label{fig:1-2}
\end{figure*}

In the traditional view, the spatial mode transformation and nonlinear frequency conversion should be two independent symmetric processes, say, the sequence of mode and frequency conversion should not affect the converted result of structured light. For instance, an HG$_{0,1}$ mode (Hermite-Gaussian mode with a unit topological index, $m=1$) can be converted to LG$_{0,1}$ mode (Laguerre-Gaussian mode with a unit OAM index, $\ell=1$) via an astigmatic mode conversion ($\hat{\textbf{M}}$), see Fig.~\ref{f1}(a), and then the topological charge can increase to two in an SHG conversion of a nonlinear crystal ($\hat{\textbf{F}}$). Under the opposite sequence of frequency and astigmatic ($\hat{\textbf{F}}$-$\hat{\textbf{M}}$) conversions, the final converted modes will have the same result of double frequency and OAM charge (sometimes the splitting of two generic singularities was considered, but it does not affect the conclusion). In this work, in contrast to the traditional view, we unveil that sequences of frequency and mode conversions are not always commutative for topological distribution. The conversions of a complex structured light $\Psi$ with hidden spatial symmetry can illustrate this rule obviously, see Fig.~\ref{f1}(b), where a strong asymmetry of $\hat{\textbf{F}}$ and $\hat{\textbf{M}}$ conversions is unveiled in converted results. Reversing the sequence of the two conversions, the spatial symmetry and topological charges of result will both dramatically change. Hereinafter, we will theoretically and experimentally demonstrate the generalized inherent (non)commutative rules in $\hat{\textbf{F}}$-$\hat{\textbf{M}}$ asymmetric nonlinear conversions of complex structured light.

The astigmatic mode conversion of structured light can be represented by a Poincar\'e sphere~\cite{dennis2017swings,alonso2017ray}, which can also extended to complex structured modes as spatial wave packets of superposted eigenmodes~\cite{shen20202,gutierrez2020modal}. Here, we combine SHG conversion with the Poincar\'e sphere to cover all generalized converted results with sequences of $\hat{\textbf {F}}$ and $\hat{\textbf {M}}$ conversions, see Fig.~\ref{fig:1-2}. The different positions on the conversion Poincar\'e sphere represent structured mode with different astigmatism, and different circles in gray, orange and green around the sphere represent FW structured light or their corresponding SHG structured light in different conversion sequences respectively. 

\begin{figure*}[htbp]
 \setlength{\abovecaptionskip}{0pt}
 \setlength{\belowcaptionskip}{-12pt}
\centering\includegraphics[width=0.82\linewidth]{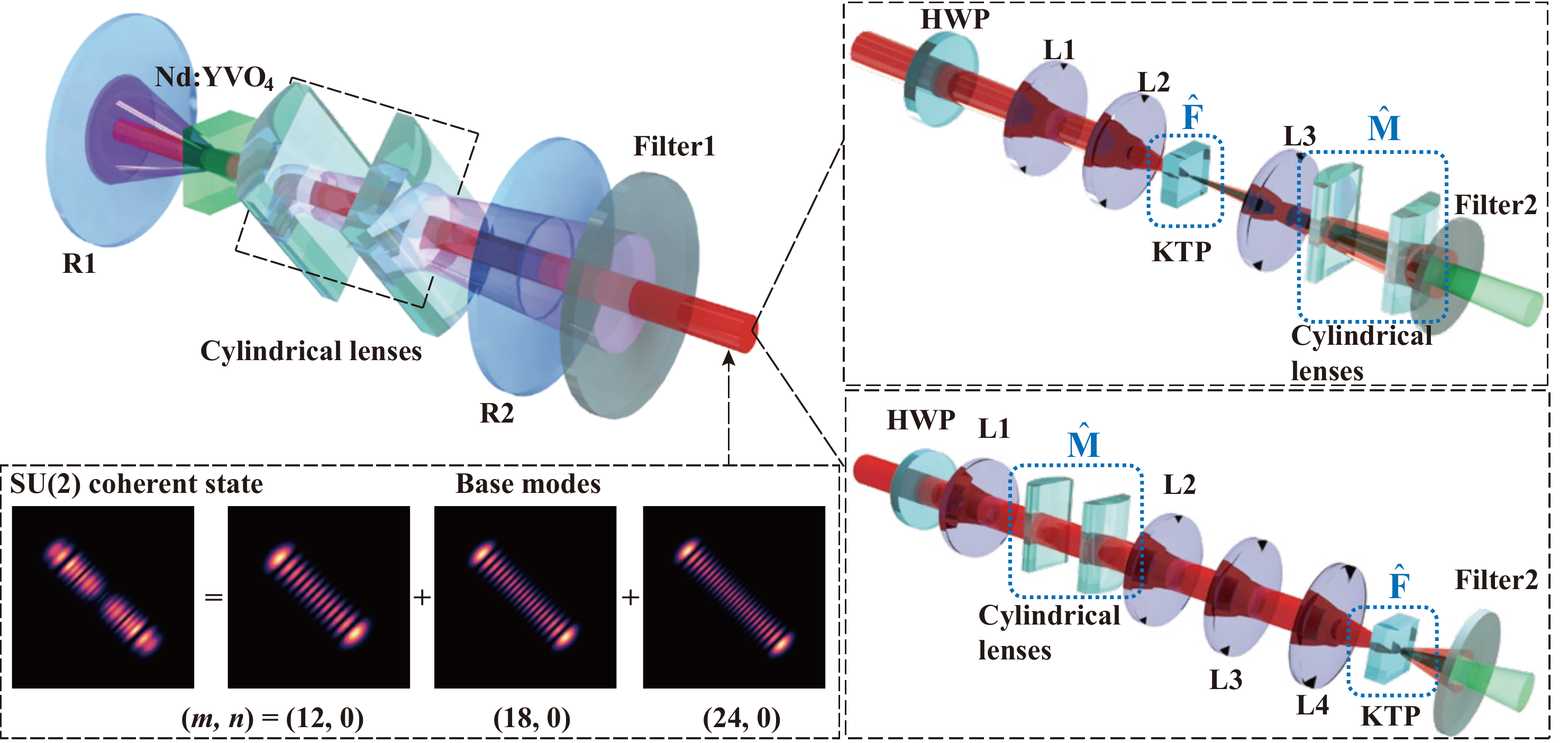}
\caption{The experimental setup. The FW beam is generated via a solid-state laser with Nd:YVO$_{4}$ as the gain medium (left). Two types of conversion processes are on the right. The left-bottom inset implies the decomposition of an SU(2) geometric mode. For the right $\hat{\textbf{F}}$-$\hat{\textbf{M}}$ part, the upper panel corresponds to the $\hat{\textbf{F}}$ first case, and the lower panel represents the $\hat{\textbf{M}}$ first case. R1: concave dichroic mirror (Anti-reflective (AR) at $808~\rm{nm}$ and high-reflective (HR) at $1064~\rm{nm}$, radius of curvature $R_{1}$: $206.3~\rm{mm}$); R2: output coupler (transmittance 40$\%$ at $1064~\rm{nm}$, radius of curvature $R_{2}$: $1000~\rm{mm}$); Filter1: AR from $1000$ to $1400~\rm{nm}$, filtering off pumping light; Half-wave plate (HWP), adjusting the polarization of generated beams for higher nonlinear conversion efficiency; Filter2: AR at $532~\rm{nm}$, filtering off $1064~\rm{nm}$ FW beam; Cylindrical lenses: focal lengths $f_{0}$=$25~\rm{mm}$; Lens group: L1, L2, L3, L4; KTP: frequency doubling crystal.} 
\label{fig:1}
\end{figure*}

To study (non)commutative rules of these two conversions, we defined two types of sequences: i) frequency first and astigmatism next where converted SHG modes are $E_{2}=\hat{\textbf{M}}\hat{\textbf {F}}E_{1}$, and ii) astigmatism first and frequency next where converted SHG modes are $E_{2}=\hat{\textbf {F}}\hat{\textbf {M}}E_{1}$, respectively. We study the structured mode evolution in the SHG process mathematically based on the coupled wave equation~\cite{zhao2020conversion}, which is rewritten as Eq.~(\ref{eq:2}) combined with the wave equation.
\begin{equation}
 \setlength{\abovedisplayskip}{0pt}
 \setlength{\belowdisplayskip}{-3pt}
		\sum_{m,n}\frac{dB_{m,n}(z)}{dz}U_{2_{m,n}}(x,y,z)=-\frac{1}{2}iK_{1}E_{1}^2(x,y,z)
		\label{eq:2}.
\end{equation}
where $K_{1}$ the coupling coefficient, and $U_{2_{m,n}}$ each eigenmode component of SHG modes with $B_{m,n}$ of their coefficients. In our general case, the FW beam $E_{1}$ is set as the SU(2) geometric mode, which is the coherent superposition of multiple frequency-degenerate eigenmodes controlled by multiple DoFs~\cite{shen2020structured,shen2021rays}, and expressed as $E_{1}(x,y,z)=\sum_{k=0}^{N}2^{-\frac{N}{2}}{\binom{N}{k}}^{1/2}\rm{e}^{\textit{ik}\myphi_{\rm{0}}}\it{U_{\rm{1}_{m,n}}(x,y,z)}$, where $U_{1_{m,n}}$ represents eigenmodes, indices $m=m_{0}+qk$, $n=n_{0}+pk$, $m_{0}$ and $n_{0}$ are initial orders, $N$ is related to the number of eigenmodes, $\myphi_{0}$ is initial phase, and $p$ and $q$ are coprimed integers (see Supplementary Materials for theoretical details). When $N$ is set as zero, FW beam $E_{1}$ is reduced to a pure eigenmode. Therefore the following derivations are correct both for single modes and for multiplexed modes.

For HG mode bases, $U_{\myxi_{m,n}}=U^{(0,0)}_{\myxi_{m,n}}$, where $\myxi=1$ or $2$ represents FW or SHG. The coefficients $B_{m,n}(z)$ as well as the light field distributions of the SHG waves can be obtained by integrating to $z$ (see Supplementary Materials for calculation details). The SHG process is represented with the operator $\hat{\textbf {F}}$ and we have $E^{(0,0)}_{2}=\hat{\textbf {F}}E^{(0,0)}_{1}$. In our study, the HG structured modes directly emitted from a laser cavity are taken as the initial modes before the astigmatic conversion. To quantitatively evaluate the results, we resort to general Hermite-Laguerre-Gaussian (HLG) modes, which can be generated based on HG modes~\cite{PhysRevA.83.032124} with different astigmatism-conversion-related parameters. The obtained HLG mode with variable $\theta$ and $\beta$ can be denoted as $U^{(\beta,\theta)}_{\myxi_{m,n}}$, where $\theta$ and $\beta$ are tunable astigmatic parameters. To simplify the calculation description, the astigmatic conversion process is represented with operator $\hat{\textbf {M}}$ thus we have $E^{(\beta,\theta)}_{\myxi_{m,n}}=\hat{\textbf {M}}E_{\myxi_{m,n}}^{\left(0,0\right)}$. 

Firstly, we take HG$_{1,0}$ as an example shown in Fig.~\ref{fig:1-2}(a). The gray circle corresponds to FW modes with different angular parameters ($\theta$, $\beta$) of astigmatic conversion, whose first and the second type converted modes are shown in the orange and green circles respectively. On the black equator line, the first and the second type converted modes have the same components with the same $B_{m,n} (z)$. That means the symmetry of a single eigenmode can be achieved with $\beta=0$, mathematically,
\begin{equation}
 \setlength{\abovedisplayskip}{0pt}
 \setlength{\belowdisplayskip}{0pt}
	\begin{split}
\hat{\textbf {M}}\hat{\textbf {F}}E_{1}-\hat{\textbf {F}}\hat{\textbf {M}}E_{1}=[\hat{\textbf {M}},\hat{\textbf {F}}]E_{1}=0.
		\label{eq:0--1}
			\end{split}
		\end{equation}
When $\beta\ne0$, modes on the Poincar\'e sphere are demonstrated to be asymmetric after two conversions, which is manifested through different $B_{m,n} (z)$ of each base mode,
\begin{equation}
 \setlength{\abovedisplayskip}{0pt}
 \setlength{\belowdisplayskip}{0pt}
	\begin{split}
\hat{\textbf {M}}\hat{\textbf {F}}E_{1}-\hat{\textbf {F}}\hat{\textbf {M}}E_{1}=[\hat{\textbf {M}},\hat{\textbf {F}}]E_{1}\neq0.
		\label{eq:0--2}
			\end{split}
		\end{equation}

As a result, intensity distributions of the first and the second type converted modes are different, as shown in Fig.~\ref{fig:1-2}(a). For the results of SU(2) geometric mode, such symmetry and asymmetry can also be obtained more intuitively from its corresponding Poincar\'e sphere as shown in Fig.~\ref{fig:1-2}(b). For instance, modes in the gray circle represents FW SU(2) geometric modes with parameters $N=2, q=3, p=0, m_{0}=1, n_{0}=0$. The orange and green circles are the corresponding first and second type converted SHG modes respectively. On the black equator line, the first and second type converted SHG modes share the same $B_{m,n} (z)$ for each mode component and share the same intensity distributions. For cases of $\beta\ne0$, as highlighted by the blue points, $\hat{\textbf{F}}$-$\hat{\textbf{M}}$ conversions are asymmetric for those SHG modes. Such asymmetry of $\hat{\textbf{F}}$-$\hat{\textbf{M}}$ conversions also means that one can obtain more nonlinear wavelength degenerate modes, which provides a new controllable dimension of converted structured light.

In experiments, we generate modes on the pole of Poincar\'e spheres with angular parameters $\theta=\pi/2$ and $\beta=\pi/2$ to demonstrate above theoretical discoveries. The setup includes the intracavity FW modes generation part and extracavity conversion part. Mirrors R1 and R2 form the laser cavity and Nd:YVO$_{4}$ is pumped by an $808~\rm{nm}$ laser diode. With off-axis of R1, HG modes with one tunable index can be generated directly from cavity. To generate 2D HG modes with two tunable indices in the transverse plane, two cylindrical lenses are introduced into the cavity to control the second index of modes through off-axis of the first cylindrical lens~\cite{pan2020index}. Furthermore, by precisely controlling distances between cavity elements, intracavity HG modes can realize so-called frequency-degeneracy, and consequently SU(2) geometric modes can be generated from cavity as the below subfigure of the left panel of Fig.~\ref{fig:1}. Then the FW beam enters mode conversion parts, including two conversions as presented before: frequency and astigmatism conversions. The $\hat{\textbf{F}}$ conversion is realized with a KTP crystal and the $\hat{\textbf{M}}$ conversion is achieved by a mode converter composed of two cylindrical lenses. In Fig.~\ref{fig:1}, the upper and lower parts of the right panel show the first and second type conversions respectively (see more experimental details in Supplementary Material).

\begin{figure}[htbp]
 \setlength{\abovecaptionskip}{-3pt} \setlength{\belowcaptionskip}{-15pt}
\centering\includegraphics[width=\linewidth]{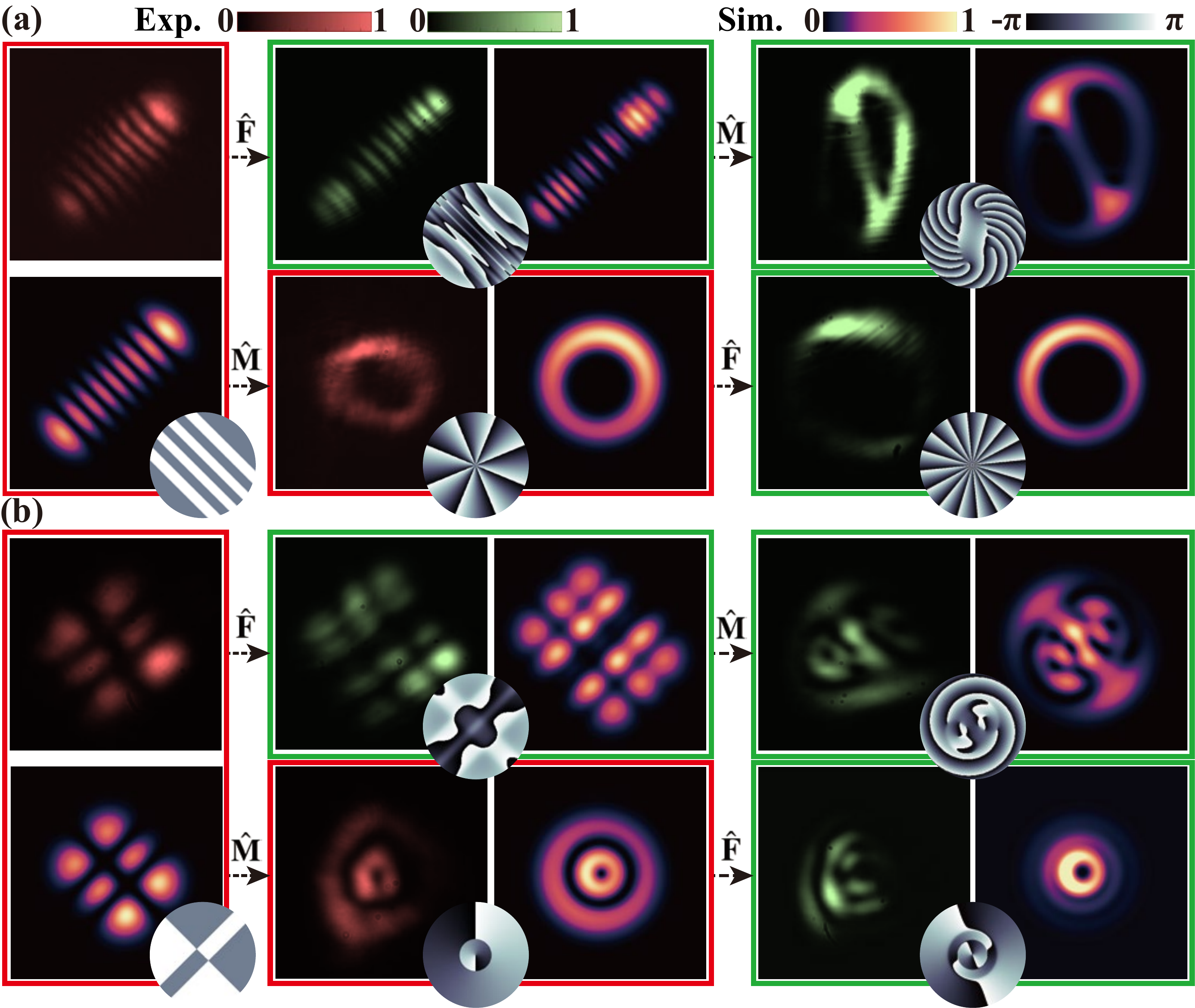}
\caption{Experimental (red and green colorbars) and simulated results about the first and second type conversions for (a) 1D HG mode with $(m, n)=(0,8)$ and (b) 2D HG mode with $(m, n)=(2, 1)$. The experiments and simulations match well in demonstrating the proposed asymmetry, including the pair at the bottom right panel of this figure, where simulated result also has three circles in radial direction. Their corresponding phase distributions are shown in the circle insets.}
\label{fig:2}
\end{figure}


\begin{figure}[htbp]
 \setlength{\abovecaptionskip}{-3pt}
 \setlength{\belowcaptionskip}{-19pt}
\centering\includegraphics[width=0.9\linewidth]{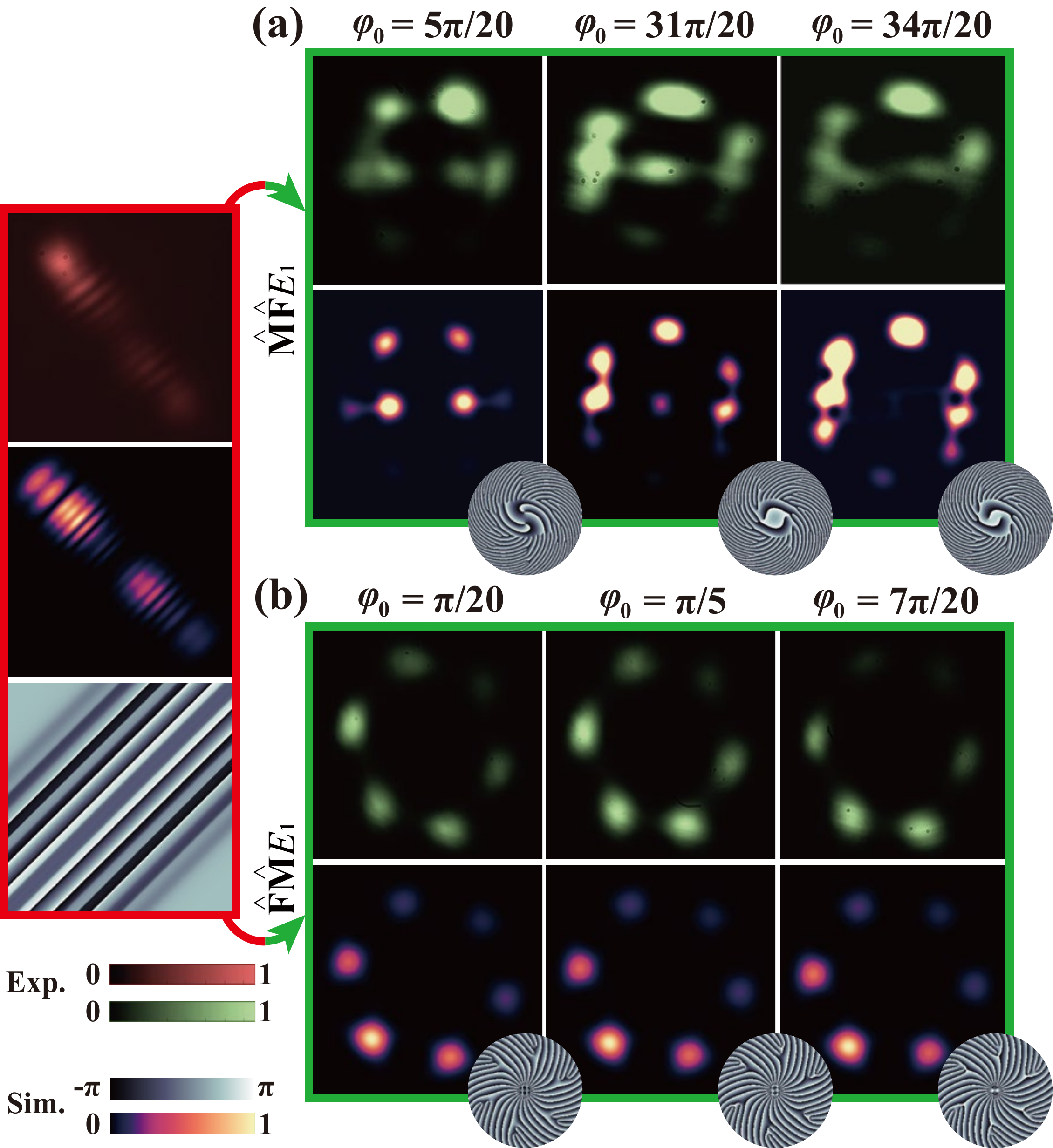}
\caption{Experimental (red and green colorbars) and simulated results of (a) the first type and (b) the second type converted SU(2) geometric modes on the polar of the conversion Poincar\'e sphere with different $\myphi_{0}$. Their corresponding phase distributions are shown in the circle insets.}
\label{fig:3}
\end{figure}

We generate $1064~\rm{nm}$ HG$_{0,8}$ mode directly as the left red frame of Fig.~\ref{fig:2}(a). When it goes through KTP crystal, corresponding $532~\rm{nm}$ SHG modes are generated as shown in the middle green frame of Fig.~\ref{fig:2}(a), which is not a single HG mode but an HG-based superposed mode. The index $n$ of SHG HG modes includes only even numbers from 0 to 16. {The experimental asymmetrical intensity distributions are caused by the intensity asymmetry of the FW beams generated from the cavity with a complex-valued shift of conventional beams in the Cartesian plane~\cite{kovalev2016asymmetric} which are also considered in simulation (see in Supplementary Material).} After the $\hat{\textbf{M}}$ conversion, every $532~\rm{nm}$ HG eigenmode component becomes a corresponding LG eigenmode component with even topological charges from $-16$ to 0. The first type converted SHG mode is shown in the upper right green frame of Fig.~\ref{fig:2}(a). In contrary, for the second type conversion, the HG FW mode is astigmatism converted first to generate a LG mode with $(p,\ell)=(0,-8)$ as shown in the middle red frame of Fig.~\ref{fig:2}(a). After $\hat{\textbf{F}}$ conversion, the corresponding SHG mode is a superposed mode with different indices of $p$ but the same $\ell=-16$ as seen in the lower right green frame of Fig.~\ref{fig:2} (a). The first type converted SHG mode is obviously different with the second type from the comparisons of the right two columns in Fig.~\ref{fig:2}(a), as well as the phase distributions of the insets, which clearly demonstrates the asymmetry of $\hat{\textbf{F}}$-$\hat{\textbf{M}}$ conversions of a single HG mode on the polar of the conversion Poincar\'e sphere. Moreover, the final mode results of 2D HG$_{2,1}$ mode after these two types of conversions are presented in Fig.~\ref{fig:2}(b). The single $1064~\rm{nm}$ HG mode becomes a SHG HG-based mode in the middle green frame of Fig.~\ref{fig:2}(b) after $\hat{\textbf{F}}$ conversion. Index $n$ includes 0 and 2 while index $m$ includes 0, 2, 4. After $\hat{\textbf{M}}$ conversion, this HG-based mode becomes a superposed LG-based mode with $\ell$ from $-2$ to 4 as displayed in the upper right green frame of Fig.~\ref{fig:2}(b). However, when the single HG mode goes through $\hat{\textbf{M}}$ conversion first, a LG mode with $(p,\ell)=(1,1)$ is generated, and after $\hat{\textbf{F}}$ conversion, corresponding SHG mode is generated as a superposed mode with different index $p$ but the same $\ell=2$ in the lower right green frame of Fig.~\ref{fig:2}(b). The phase distributions under these two different conversion sequences are shown in the insets respectively.

Based on the designed distances between cavity elements, we then generated the SU(2) geometric mode with HG bases directly (see the red frame of Fig.~\ref{fig:3}). The SU(2) geometric mode with $N=2$, $(q, p)=(6, 0)$ and $(m_{0}, n_{0})=(12, 0)$ composes of three HG bases. As this $1064~\rm{nm}$ SU(2) geometric mode evolves in the first type conversion on the polar of the SU(2) Poincar\'e sphere, the $532~\rm{nm}$ LG-based SU(2) geometric modes are finally shown in Fig.~\ref{fig:3}(a). With little disturbance of the spatial position of KTP, the initial phase $\myphi_{0}$ of the SHG modes will change, and the $532~\rm{nm}$ LG-based SU(2) geometric mode also varies with different $\myphi_{0}$ as reflected by both the rotating angle and intensity pattern. If the $1064~\rm{nm}$ HG-based mode is astigmatism $\hat{\textbf{M}}$-converted first and then $\hat{\textbf{F}}$-converted, the $532~\rm{nm}$ LG-based SU(2) geometric modes are shown in Fig.~\ref{fig:3}(b). Different from that in (a), with different $\myphi_{0}$, the $532~\rm{nm}$ LG-based SU(2) geometric mode evolves only on the rotating angle without global intensity distribution. From phase distributions in insets of Fig.~\ref{fig:3}, the OAM distributions have obvious differences between two type converted results, especially in the central parts. In short, this part illustrates the asymmetry in $\hat{\textbf{F}}$-$\hat{\textbf{M}}$ of SU(2) geometric modes on the polar of the conversion Poincar\'e sphere, and provides a new dimension to generate more tunable wavelength-degenerate modes.
In conclusion, we consider topological distribution evolution in the SHG process caused by both frequency-doubling and astigmatic mode conversion, and unveil the (non)commutative
rule of the $\hat{\textbf{F}}$-$\hat{\textbf{M}}$ sequence on the SHG structured beams. Different from the traditional view, only astigmatic index $\beta=0$ corresponds to the commutative $\hat{\textbf{F}}$-$\hat{\textbf{M}}$ conversion sequences, and other astigmatic modes on the Poincar\'e sphere are non-commutative for $\hat{\textbf{F}}$-$\hat{\textbf{M}}$ conversion sequences. Our results are suitable for both single eigenmodes and complex structured light with multiple
DoFs. This effect not only creates extended structured laser modes but also serves as new
physics in nonlinear light field manipulation highlighting important applications in imaging, optical trapping, communications, entangled photons generation, and nonreciprocal devices broadly.

\bibliographystyle{naturemag}

\textbf{Acknowledgments.} National Natural Science Foundation of China (62275137, 61975087).

\bibliography{sample}

\end{document}